\documentclass[reprint]{revtex4-1}

\usepackage{graphicx}
\usepackage{bm}
\usepackage{amsmath}
\usepackage{amssymb}

\begin{document}


\title{Spin-spin correlation function of the $2D$ $XY$ model with weak site or bond dilution}


\author{Oleksandr Kapikranian}
\email{akap@icmp.lviv.ua}
\affiliation{Institute for Condensed Matter Physics, National Academy of
             Sciences of Ukraine, 79011 Lviv, Ukraine}

\date{\today}

\begin{abstract}
The spin-spin correlation function of the $2D$ $XY$ model decays as a power law at all temperatures below the Berezinskii-Kosterlitz-Thouless transition point with a temperature dependent exponent $\eta=\eta(T/J)$ ($J$ is the ferromagnetic coupling strength). It is known from computer experiments that in the $2D$ $XY$ model with site or bond dilution this exponent depends on concentration $p$ of removed sites/bonds as well. Knowing the slope $\partial\eta/\partial p$ at point $p=0$, one can predict the value of the exponent for small dilution concentrations: $\eta(p)\simeq\eta(0)+p(\partial\eta/\partial p)|_{p=0}$. As it is shown in this paper, the spin-wave Hamiltonian allows to obtain exact results for this slope: $(\partial\eta/\partial p)|_{p=0} = T/(2J) + O((T/J)^2)$ and $T/(\pi J) + O((T/J)^2)$ for site and for bond dilution, respectively.
\end{abstract}

\pacs{05.50.+q; 75.10.Hk}

\maketitle

\section{\label{intro}Introduction}

An effect produced by introduction of structural randomness is perhaps one of the first aspects one would be willing to investigate, once the properties of the model of interest have been successfully studied on regular structures. While computer experiment data keep accumulating for diverse models with structural disorder, this problem is often a real challenge to the theory, though.

We consider the two-dimensional $XY$ model (sometimes referred to as the planar rotator model), which Hamiltonian is traditionally written as 
\begin{equation}\label{H_2dXY}
H = - J \sum_{\left<{\bf r,r'}\right>} \cos(\theta_{\bf r} - \theta_{\bf r'})
\end{equation}
with the sum spanning the pairs of nearest neighbors in a square lattice of $N$ sites, $J>0$ being the coupling strength, and the polar angle $\theta_{\bf r}$ representing the only degree of freedom which can be attributed to a spin of unit length rotating in a plane.

The $2D$ $XY$ model is remarkable for its critical properties, as this particular combination of lattice dimensionality and spin symmetry leads to the existence of a finite range of temperatures in which the system exhibits critical-like behaviour [Berezinskii-Kosterlitz-Thouless (BKT) phase] \cite{KosterlitzThouless73, Berezinskii72}; most notably, the spin-spin correlation function decays as a power law with a temperature dependent exponent $\eta=\eta(T/J)$ below the BKT transition point $T_{\mathrm{BKT}}$.

In the low temperature limit, where the spin-wave approximation (SWA) is applicable, i. e. the cosine in the Hamiltonian (1) can be replaced by a quadratic expression without affecting the system properties significantly, one arrives easily at a power law form of the spin-spin correlation function, $R^{-\eta}$, with an exponent linearly dependent on temperature \cite{Rice65,Wegner67}: 
\begin{equation}\label{eta_swa}
\eta_{\mathrm{SWA}}=T/2\pi J .
\end{equation}
It is known, however, that as the temperature increases, the real exponent increases non-linearly with temperature, so that it assumes the exact value of $1/4$ at $T_{\mathrm{BKT}}$ \cite{Kosterlitz74}.

Given the two both theoretically and experimentally (computer experiment is meant here) acknowledged facts that the value of the exponent $\eta$ at the BKT transition point cannot be changed by structural dilution (see, for example, \cite{SurunganOkabe05}) whereas the value of the BKT transition temperature is reduced by dilution and depends on its concentration \cite{SurunganOkabe05,KapikranianEtAl08,WysinEtAl05}, one can already make a conclusion that the value of $\eta$ below the BKT point should depend not only on temperature but on dilution concentration as well. It is also clear that $\eta$ should increase with dilution concentration for $T<T_{\mathrm{BKT}}$. It can be interpreted as the increase of effective temperature (decrease of effective interaction) due to dilution..

A number of works have touched this question, mostly using computer simulations. For site dilution, when some fraction of sites is excluded from Hamiltonian (\ref{H_2dXY}), see \cite{WysinEtAl05, BercheEtAl03, KapikranianEtAl07, SunEtAl09}, and for bond dilution, when some fraction of bonds is removed from (\ref{H_2dXY}), see \cite{SurunganOkabe05}. 

The present study logically continues the theoretical part of \cite{KapikranianEtAl07}, making a significant advance~\cite{footnote} and covering both the site and bond dilution cases. The focus is on the behavior of the spin-spin correlation function and the searched quantity is the dilution concentration $p$ dependent exponent $\eta$ of the correlation function power-law decay. It is natural to assume that the exponent $\eta=\eta(T/J,p)$ is an analytic function with respect to $p$, away from the percolation threshold. Below, $p$ will denote the fraction of removed bonds or sites, depending on what dilution type is considered. Thus, $\eta$ can be presented as a power series 
\begin{equation}\label{eta_series}
\eta(p)\simeq\eta(0)+p(\partial\eta/\partial p)|_{p=0}+\cdots.
\end{equation}
For small dilution concentrations $p$, it is enough to know the slope $(\partial\eta/\partial p)|_{p=0}$ to estimate the value of exponent $\eta$ with good precision. So, in our derivation we drop terms that lead to higher order terms in $p$ in (\ref{eta_series}).

As more simple and transparent from the technical point of view case, bond dilution is considered first in Section \ref{II}, where the spin-spin correlation function is calculated up to the contributions linear in dilution concentration $p$ and temperature $T/J$. The analogous but more technically involved derivation for the correlation function of a system with site dilution can be found in Section \ref{III}. The final results for the exponent of the spin-spin correlation function of the systems with site and bond dilution are given, respectively, by Eqs. (\ref{eta_s.d.}) and (\ref{eta_b.d.}) (see Fig. \ref{fig2}).

\section{\label{II} $2D$ $XY$ model with bond dilution}

In this section the case of bond dilution in the $2D$ $XY$ model is considered.
First, in Subsection \ref{II-1}, the bond diluted spin-wave Hamiltonian and the procedure of configurational averaging are defined. Then, in Subsection \ref{II-2}, the spin-spin correlation function is calculated up to the contributions linear in dilution concentration $p$ and temperature.


\subsection{\label{II-1} Bond diluted Hamiltonian and configurational averaging}\label{B.d. Hamilt. and conf.avrg.}

\begin{figure}[h]
\center{\includegraphics[width=0.35\textwidth,angle=0]{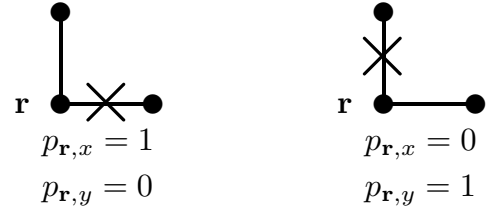}}
\caption{\label{fig1} The occupation number $p_{{\bf r},\alpha}$ ($\alpha = x,y$) takes value $1$ if bond $({\bf r},{\bf u}_\alpha)$ is removed and $0$ otherwise.}
\end{figure}

Hamiltonian (\ref{H_2dXY}) in the SWA and with bond dilution can be written as
\begin{equation}\label{H_b.d.}
H_{\mathrm{b.d.}} = \frac{J}{2} \sum_{\bf r}\sum_{\alpha=x,y} (\theta_{\bf r} - \theta_{{\bf r}+{\bf u}_\alpha})^2 (1 - p_{{\bf r},\alpha}),
\end{equation}
where ${\bf u}_x = (a,0)$, ${\bf u}_y = (0,a)$ ($a$ is the lattice spacing), and $p_{{\bf r},\alpha}=1$ if bond $({\bf r,u}_\alpha)$ is removed and $0$ otherwise (see Fig. \ref{fig1}). Then, any thermodynamic quantity characterizing the system will depend on the particular choice of configuration $\{p_{{\bf r},\alpha}\}$ of the discrete variables. 

One is willing to consider here what is often referred to as quenched dilution, i. e. when there is a fixed fraction $p$ of removed bonds distributed randomly in the system and frozen at their position \cite{Brout59}. Meaningful physical quantities can be obtained averaging them over the configurations with a fixed fraction of removed bonds $p$. For a large system one might as well allow \emph{all} configurations, ascribing them a probabilistic weight
\begin{eqnarray}
P(\{p_{{\bf r},\alpha}\}) = \prod_{{\bf r},\alpha} \left[(1-p)(1-p_{{\bf r},\alpha}) + p p_{{\bf r},\alpha}\right]\nonumber\\ = (1-p)^{\sum_{{\bf r},\alpha}(1-p_{{\bf r},\alpha})}p^{\sum_{{\bf r},\alpha}p_{{\bf r},\alpha}},
\end{eqnarray}
meaning that a bond is removed with probability $p$, which will lead to the fact that only realizations with fraction $\sum_{{\bf r},\alpha}p_{{\bf r},\alpha}/(2N) \simeq p$ ($2N$ is the number of bonds in the initial lattice) of removed bonds will make essential contribution to the averaged quantities, when $N\to\infty$. It immediately follows that
\begin{equation}\label{p^i_avrg}
\overline{p^i_{{\bf r},\alpha}} = p,\quad \overline{p_{{\bf r}_1,\alpha_1} \cdots p_{{\bf r}_i,\alpha_i}} = p^i
\end{equation}
(all pairs $({\bf r}_1,\alpha_1), \ldots, ({\bf r}_i,\alpha_i)$ are different), where $\overline{(\ldots)}$ means averaging with respect to disorder configurations, 
$$
\overline{(\ldots)} = \left(\prod_{{\bf r},\alpha}\sum_{p_{{\bf r},\alpha}=0,1}\right) P(\{p_{{\bf r},\alpha}\}) \ldots ,
$$ 
hereafter referred to as configurational averaging.

It is convenient to rewrite Hamiltonian (\ref{H_b.d.}) in the Fourier transformed variables $\theta_{\bf k} = \frac{1}{\sqrt{N}}\sum_{\bf r}e^{i{\bf kr}}\theta_{\bf r}$ as
\begin{equation}\label{H_b.d._four}
H_{\mathrm{b.d.}} = H_0 + H(\{p_{{\bf r},\alpha}\}), \ H(\{p_{{\bf r},\alpha}\})\equiv\sum_{{\bf r},\alpha}p_{{\bf r},\alpha} H_\alpha({\bf r}),
\end{equation}
where
\begin{equation}\label{H_pure_four}
H_0 = -J\sum_{{\bf k}}\gamma_{\bf k}\theta_{\bf k}\theta_{\bf -k}
\end{equation}
with 
\begin{equation}\label{gamma}
\gamma_{\bf k} = 2\left(\sin^2\frac{k_xa}{2} + \sin^2\frac{k_xa}{2}\right)
\end{equation}
is the Hamiltonian of the undiluted system, and
\begin{equation}\label{H_alpha(r)}
H_\alpha({\bf r}) = -\frac{J}{2N}\left[\sum_{{\bf k}}e^{-i{\bf kr}}(1-e^{-ik_\alpha a})\theta_{\bf k}\right]^2 .
\end{equation}
The sums over ${\bf k}$ in (\ref{H_pure_four}) and (\ref{H_alpha(r)}) span the 1st Brillouin zone.

The thermodynamic average of some physical quantity $A$ can be written as
\begin{equation}\label{<A>}
\left<A\right> = \textrm{Tr}_\theta Ae^{-\beta H_{\mathrm{b.d.}}}/\textrm{Tr}_\theta e^{-\beta H_{\mathrm{b.d.}}}
\end{equation}
Since $\theta_{\bf k}$ is a complex variable (for ${\bf k} \neq 0$): $\theta_{\bf k} = \theta^c_{\bf k} + i \theta^s_{\bf k}$, $\mathrm{Tr}_{\theta}$ above means
\begin{equation}\label{Tr_phi}
\mathrm{Tr}_{\theta} = \int d\theta_0 \prod_{{\bf k}\in
B/2}\int_{-\infty}^\infty d\theta^c_{\bf k}
\int_{-\infty}^\infty d\theta^s_{\bf k}\ ,
\end{equation}
where $B/2$ stands for a half of the 1st Brillouin zone excluding ${\bf k} = 0$ ($\theta^c_{\bf k}$ and $\theta^s_{\bf k}$ in the other half are not independent due to the relations: $\theta^c_{\bf -k} = \theta^c_{\bf k}$ and $\theta^s_{\bf -k} = -\theta^s_{\bf k}$). Note, that it is possible to extend the bounds of integration in (\ref{Tr_phi}) to infinity, since the functions that stand after the trace are always rapidly decaying at $\beta J\to\infty$.

The configurationally averaged value of $\left<A\right>$ can be obtained using the Taylor series representations of the exponential and $(1+x)^{-1}$ functions with respect to powers of $H(\{p_{{\bf r},\alpha}\})$. The equalities in (\ref{p^i_avrg}) easily lead to 
\begin{widetext}
\begin{eqnarray}\nonumber
&&\overline{H^i(\{p_{{\bf r},\alpha}\})} = p\sum_{{\bf r},\alpha} H^i_\alpha({\bf r})
+p^2 \Bigg[\sum_{{\bf r},\alpha} \sum_{{\bf r}',\alpha'}\Bigg]' \frac{i!}{2}\sum_{i'=1}^{i-1} \frac{H_\alpha^{i-i'}({\bf r}) H_{\alpha'}^{i'}({\bf r}')}{(i-i')!i'!} + \cdots \\\label{expansion_in_p} &&
+p^n \Bigg[\sum_{{\bf r}_1,\alpha_1}\cdots \sum_{{\bf r}_n,\alpha_n}\Bigg]'\frac{i!}{n!}\sum_{i_1=1}^{i-1}\sum_{i_2=1}^{i_1-1} \cdots \sum_{i_{n-1}=1}^{i_{n-2}-1}
\frac{H_{\alpha_1}^{i-i_1}({\bf r}_1) H_{\alpha_2}^{i_1-i_2}({\bf r}_2)\cdots H_{\alpha_{n-1}}^{i_{n-2}-i_{n-1}}({\bf r}_{n-1}) H_{\alpha_n}^{i_{n-1}}({\bf r}_n)}{(i-i_1)!(i_1-i_2)!\cdots (i_{n-2}-i_{n-1})!i_{n-1}!} + \cdots ,
\end{eqnarray}
\end{widetext}
where $[...]'$ means that the terms having any coinciding pairs of indexes, ${\bf r}_i = {\bf r}_j, \alpha_i = \alpha_j$, are excluded from the sums enclosed in brackets. This result will be applied in the next subsection to calculate the spin-spin correlation function.

\subsection{\label{II-2} Spin-spin correlation function of the bond diluted $2D$ $XY$ model}

The spin-spin correlation function of the $XY$ model described by Hamiltonian $H$ can be written as
\begin{equation}\label{G(R)_def}
G({\bf R}) = \Re \left<e^{i(\theta_{\bf R} - \theta_0)}\right> =
\Re \frac{\mathrm{Tr}_\theta e^{-\beta H + i\sum_{\bf k}\eta_{\bf
k}({\bf R})\theta_{\bf k}}}{\mathrm{Tr}_\theta e^{-\beta H}}
\end{equation}
with
\begin{eqnarray}\label{eta_def}
\eta_{\bf k} ({\bf R}) &=& \left( e^{-i\bf kR} - 1
\right)/\sqrt{N} .
\end{eqnarray}
For the undiluted system, Eq. (\ref{H_pure_four}), one can write, since $\theta^c_{\bf -k} = \theta^c_{\bf k}$ and $\theta^s_{\bf -k} = -\theta^s_{\bf k}$, using the notations of (\ref{Tr_phi}), 
\begin{eqnarray}\nonumber
&&G_0({\bf R}) = \Re \mathrm{Tr}_\theta e^{-2\beta J\sum_{{\bf k}\in
B/2}\gamma_{\bf k}\left[(\theta^c_{\bf k})^2 + (\theta^s_{\bf k})^2\right]}
\\\nonumber
&&\times e^{2i\sum_{{\bf k}\in B/2}(\eta^c_{\bf k}\theta^c_{\bf k} - \eta^s_{\bf k}\theta^s_{\bf k})} / \mathrm{Tr}_\theta e^{-2\beta J\sum_{{\bf k}\in
B/2}\gamma_{\bf k}\left[(\theta^c_{\bf k})^2 + (\theta^s_{\bf k})^2\right]},
\\\label{G(R)_four}
\end{eqnarray}
where $\eta^c_{\bf k}$ and $\eta^s_{\bf k}$ denote the real and imaginary parts of $\eta_{\bf k} ({\bf R})$. It is straightforward to get from the Gaussian integration:
\begin{equation}\label{G0(R)}
G_0({\bf R}) =  \exp\left[-\frac{1}{4\beta J}\sum_{{\bf k}\neq
0} \eta_{\bf k} ({\bf R}) \eta_{\bf -k} ({\bf R})/ \gamma_{\bf k}\right],
\end{equation}
here and below sums over $\bf k$ span the entire 1st Brillouin zone except the point ${\bf k} = 0$.

To obtain the asymptotic behaviour of (\ref{G0(R)}) at $R\to\infty$ one should use the fact that $\eta_{\bf k} \eta_{\bf -k} = \frac{4}{N}\sin^2\frac{\bf kR}{2}$ oscillates very fast comparing to $1/\gamma_{\bf k}$ and, thus, can be replaced by its average value $2/N$ everywhere expect the region close to the singularity point ${\bf k}=0$. In this region, replacing in the thermodynamic limit $N\to\infty$ the sum with an integral and taking the leading terms of the Taylor expansion of $\sin^2\frac{\bf kR}{2}$ and $\gamma_{\bf k}$, one gets an integrable expression. One arrives at (see, for example, \cite{Wegner67} or \cite{KapikranianEtAl07} for details)
\begin{equation}\label{sum_etaeta/gamma}
\sum_{{\bf k}\neq
0} \eta_{\bf k} ({\bf R}) \eta_{\bf -k} ({\bf R})/ \gamma_{\bf k} \underset{R\to\infty}{\rightarrow} \frac{2}{\pi}\ln\frac{R}{a} + \mathrm{const}.
\end{equation}
It is easy to see that this asymptotic expression leads to a power-law decay of the spin-spin correlation function, $R^{-\eta}$, with an exponent given by (\ref{eta_swa}).

For a system with bond dilution the spin-spin correlation function is given by (\ref{G(R)_def}) with $H=H_{\mathrm{b.d.}}$, Eqs. (\ref{H_b.d._four})-(\ref{H_alpha(r)}). Applying the scheme of configurational averaging described in Subsection \ref{B.d. Hamilt. and conf.avrg.} to the correlation function, one is able to collect the resulting series into the following expression:
\begin{eqnarray}\nonumber
\overline{G({\bf R})} &=& G_0({\bf R})\Bigg\{1 + p \sum_{{\bf r},\alpha}
\Bigg(\left<e^{i\sum_{\bf k} \eta_{\bf k} \theta_{\bf
k}} e^{-\beta H_\alpha ({\bf r})}\right>_0
\\\label{overline_G_b.d.}
&&\times G^{-1}_0({\bf R})\left<e^{-\beta H_\alpha ({\bf r})}\right>^{-1}_0 - 1 \Bigg) + O(p^2)\Bigg\},\quad
\end{eqnarray}
where the terms of higher order in $p$ are dropped and $\left<\ldots\right>_0$ denotes thermodynamic averaging with Hamiltonian (\ref{H_pure_four}) of the undiluted system:
\begin{equation}\label{<...>_0}
\left<\ldots\right>_0 = \textrm{Tr}_\theta e^{-\beta H_0}\ldots / \textrm{Tr}_\theta e^{-\beta H_0}.
\end{equation}

Now, using the Taylor series representation of an exponential and the results of Appendix \ref{AppendxA} [Eqs. (\ref{theta...theta_0}) and (\ref{empty_sum_1})], one obtains for $H_\alpha ({\bf r})$ given by (\ref{H_alpha(r)})
\begin{equation}\label{avrg1}
\left<e^{-\beta H_\alpha({\bf r})}\right>_0 = 1 + \sum_{n=1}^{\infty} \frac{(2n-1)!!}{(2n)!!} \left(\frac{1}{2}\right)^n = \sqrt{2}
\end{equation}
(here and below $(2m)!!\equiv \prod_{i=1}^{m} 2i$, $(2m-1)!!\equiv \prod_{i=1}^{m} (2i-1)$, $m=1,2,\ldots$, and $0!!\equiv1$).

In a similar way, using (\ref{theta...theta_e^eta_0}), (\ref{empty_sum_1}), and the notation 
\begin{equation}
I_\alpha({\bf r}) \equiv \frac{1}{\sqrt{N}} \sum_{\bf k} e^{-i{\bf kr}}\left(1 - e^{-ik_\alpha a}\right)\eta_{-\bf k}/\gamma_{\bf k} ,
\end{equation}
one arrives at
\begin{eqnarray}\nonumber
\left<e^{i\sum_{\bf k} \eta_{\bf k} \theta_{\bf k}}e^{-\beta H_\alpha({\bf r})}\right>_0 = G_0({\bf R})\Bigg\{ 1 + \sum_{n=1}^{\infty}\sum_{l=0}^n \frac{(-1)^{n-l}}{(2\beta J)^{n-l}} 
\\\label{avrg2a}
\times \frac{(2n-1)!!}{(2l)!!(2n-2l)!} \left(\frac{1}{2}\right)^n I_\alpha^{2(n-l)}({\bf r})\Bigg\} .\qquad
\end{eqnarray}
The unity and the term with $l=n$ in (\ref{avrg2a}) give $\sqrt{2}$ [see (\ref{avrg1})]. Changing index $n\to i=n-l$ and rearranging the terms of the infinite series, one has
\begin{eqnarray}\nonumber
&&\left<e^{i\sum_{\bf k} \eta_{\bf k} \theta_{\bf k}}e^{-\beta H_\alpha({\bf r})}\right>_0 = G_0({\bf R})\Bigg\{ \sqrt{2} + \sum_{i=1}^{\infty} \frac{(-1)^{i}}{(4\beta J)^{i}} 
\\\label{avrg2b}
&&\times  \frac{I^{2i}_\alpha({\bf r})}{(2i)!} \sum_{l=0}^{\infty} \frac{(2(l+i)-1)!!}{(2l)!! 2^l} \Bigg\} = G_0({\bf R})\sqrt{2} e^{- \frac{I_\alpha^2({\bf r})}{4\beta J}} .\quad
\end{eqnarray}
The Taylor series representation of $(1 - x)^{-n/2}$,
\begin{equation}
(1 - x)^{-n/2} = 1 + \sum_{l=1}^{\infty} \frac{(2l-2+n)!!}{(2l)!!} \frac{n}{n!!} x^l,
\end{equation}
with $x=1/2$ and $n=1$ and $2i+1$ was used in (\ref{avrg1}) and (\ref{avrg2b}), respectively.

Now, having (\ref{avrg1}) and (\ref{avrg2b}), one can write the spin-spin correlation function in the low temperature limit as
\begin{eqnarray}\nonumber
\overline{G({\bf R})} &=& G_0({\bf R})\Bigg\{1 - \frac{p}{4\beta J} \sum_{{\bf r},\alpha}I_\alpha^2({\bf r}) \Bigg\} 
\\\label{b.d.corr.funct.}
&\simeq & G_0({\bf R}) e^{- \frac{p}{4\beta J} \sum_{{\bf r},\alpha}I_\alpha^2({\bf r})}.
\end{eqnarray}
Noticing that $\sum_{{\bf r},\alpha}I_\alpha^2({\bf r}) = 2\sum_{\bf k} \eta_{\bf k} \eta_{-\bf k}/\gamma_{\bf k}$, from (\ref{sum_etaeta/gamma}) immediately follows a power law decay of the correlation function, $R^{-\eta}$, with a dilution concentration dependent exponent
\begin{eqnarray}\nonumber
\eta_{\mathrm{b.d.}}(p) &=& \eta(0) +p\frac{T}{\pi J} + O(p^2) + O((T/J)^2) 
\\\label{eta_b.d.}
&\simeq& \eta(0)\left(1 + 2 p\right),
\end{eqnarray}
where $\eta(0)$ is the exponent of the pure system, Eq. (\ref{eta_swa}).

\section{\label{III} $2D$ $XY$ model with site dilution}

In this section the case of site dilution in the $2D$ $XY$ model is considered.
In Subsection \ref{III-1}, the site diluted spin-wave Hamiltonian is defined, then, in Subsection \ref{III-2}, the spin-spin correlation function is calculated up to the contributions linear in dilution concentration $p$ and temperature.

\subsection{\label{III-1}Hamiltonian of the $2D$ $XY$ model with spin
vacancies}

The spin-wave Hamiltonian of a system with site dilution differs from that of bond dilution, Eq. (\ref{H_b.d.}), in the way that the {\it four} bonds adjacent to each spinless site must be removed, so the occupation number
\begin{equation}
p_{\bf r} = \Big\{\begin{array}{ll}1,\ \text{if there is no spin on site}\ {\bf r};\\0,\ \text{otherwise,}\end{array}
\end{equation}
has to be introduced; then,
\begin{equation}\label{H0+H1}
H_{\text{s.d.}} = H_0 + H(\{p_{\bf r}\}), \quad  H(\{p_{\bf r}\}) = \sum_{\bf r} p_{\bf r} H_1({\bf r}),
\end{equation}
where $H_0$ is the Hamiltonian of the pure model, Eq. (\ref{H_pure_four}), and
\begin{equation}\label{H_1(r)}
H_1({\bf r}) = - \frac{J}{2} \sum_{\bf u} (\theta_{\bf r} -
\theta_{\bf r+u})^2
\end{equation}
with ${\bf u} = (\pm a,0), (0,\pm a)$, which in the Fourier variables reads as
\begin{equation}\label{H_1(r)_fourier}
H_1({\bf r}) = \frac{J}{N}\sum_{\bf k,k'} e^{-i{\bf (k+k')r}} g_{\bf k,k'} \theta_{\bf k} \theta_{\bf k'}
\end{equation}
with $g_{\bf k,k'} = \gamma_{\bf k,k'} - \gamma_{\bf k} - \gamma_{\bf k'}$ [$\gamma_{\bf k}$ was defined in (\ref{gamma})].

One can notice that expression (\ref{H0+H1}) is not precise when there are neighboring spin vacancies; in this case, the common bond between the vacant sites is subtracted from the ``pure" Hamiltonian twice, so it is, in fact, brought back with an opposite sign. The precise form of $H(\{p_{\bf r}\})$ would be 
\begin{equation}\label{H0+H1+H2}
H(\{p_{\bf r}\}) = \sum_{\bf r} p_{\bf r} H_1({\bf r}) + \sum_{\left<{\bf r,r}'\right>}p_{\bf r}p_{\bf r'}H_2({\bf r,r}'),
\end{equation}
where $H_2({\bf r,r'}) = \frac{J}{2} (\theta_{\bf r} - \theta_{\bf r'})^2$. However, it is not only that the second term in (\ref{H0+H1+H2}) gives contributions of order of $p^2$ and higher, after configurational averaging, but it can be {\it always} dropped when considering the spin-spin correlation function, since any non-physical extra bonds corresponding to neighboring spinless sites in (\ref{H0+H1}) are isolated from the rest of the system.

\subsection{\label{III-2} Spin-spin correlation function of the site diluted $2D$ $XY$ model}

Now, everything said in Section \ref{B.d. Hamilt. and conf.avrg.} about the bond dilution and configurational averaging can be applied to site dilution as well with the only difference that here occupation numbers $p_{\bf r}$ are defined for each site ${\bf r}$, and $p = \overline{p_{\bf r}} \simeq \sum_{\bf r}p_{\bf r}/N$ is now the fraction (concentration) of removed {\it sites}.

Then, dropping the higher order terms with respect to dilution concentration $p$, the configurationally averaged correlation function can be written as
\begin{eqnarray}\nonumber
\overline{G({\bf R})} = G_0({\bf R})\Bigg\{1 + p \sum_{\bf r}
\Bigg(\left<e^{i\sum_{\bf k} \eta_{\bf k} \theta_{\bf
k}} e^{-\beta H_1({\bf r})}\right>_0 
\\\label{G2(R)_avrg}
\times G^{-1}_0({\bf R})\left<e^{-\beta H_1({\bf r})}\right>^{-1}_0 - 1 \Bigg) +O(p^2)\Bigg\} \
\end{eqnarray}
with $\eta_{\bf k}$ given by (\ref{eta_def}).

The thermodynamic averages in (\ref{G2(R)_avrg}) can be calculated
using the Taylor series expansion: $e^{-\beta H_1({\bf r})} = \sum_{n=0}^{\infty} (-\beta H_1({\bf r}))^n/n!$. Then, the problem reduces to the calculation of the quantity $\left< e^{i\sum_{\bf k} \eta_{\bf k} \theta_{\bf k}} H^n_1({\bf r}) \right>_0$ with $\eta_{\bf k}$ given by (\ref{eta_def}) and $\eta_{\bf k} = 0$, which is presented in Appendix \ref{AppendxB}. Looking at the results (\ref{exp_1}) and (\ref{exp_2}), it is easy to see that
\begin{eqnarray}\nonumber
\left<e^{-\beta H_1({\bf r})}\right>_0 = \prod_{i=1}^{\infty}\sum_{l=0}^{\infty} \frac{1}{l!} \left((-1)^i\frac{I_i}{2i}\right)^l 
= \exp\sum_{i=1}^{\infty}(-1)^i \frac{I_i}{2i},
\end{eqnarray}
and, similarly:
\begin{eqnarray}\nonumber
\left<e^{i\sum_{\bf k} \eta_{\bf k} \theta_{\bf k}}e^{-\beta H_1({\bf r})}\right>_0 &=& G_0({\bf R}) \exp \sum_{i=1}^{\infty}(-1)^i \frac{
I_i}{2i}
\\\nonumber
&&\times \exp\left[{-\frac{1}{4\beta J}\sum_{j=1}^{\infty}(-1)^j I^*_j}\right].
\end{eqnarray}
Explicit expressions for the quantities $I_i$ and $I^*_i$ are given in (\ref{I_i_def})-(\ref{tilde_I_i}).

Finally, from (\ref{G2(R)_avrg}),
\begin{equation}\label{G2(R)_2}
\overline{G({\bf R})} = G_0({\bf R})\left\{1 + p \sum_{\bf r}
\left(e^{-{\frac{1}{4\beta J}\sum_{j=1}^{\infty}(-1)^j I^*_j}} - 1
\right)\right\}.
\end{equation}
Using the result of Appendix \ref{AppendxC}, Eq. (\ref{sum_I*_i}), with $\eta_{\bf k}$ given by (\ref{eta_def}), one has
\begin{eqnarray}\nonumber
\overline{G({\bf R})} = G_0({\bf R})\Bigg\{1 -2p + p \sum_{{\bf
r}\neq 0,{\bf R}} \Big(e^{- \frac{\pi}{8\beta J}
F_1({\bf r},{\bf R})}
\\\label{G2(R)_3}
\times e^{- \frac{\pi}{8\beta J(\pi-2)} F_2({\bf r},{\bf R})}
- 1 \Big)\Bigg\},
\end{eqnarray}
where
\begin{eqnarray}\nonumber
F_i({\bf r},{\bf R}) &=& \left[S_i(x-X,y-Y) - S_i(x,y)\right]^2 
\\\label{F_i}
&&+ \left[S_i(y-Y,x-X) -S_i(y,x)\right]^2
\end{eqnarray}
($ i=1,2$) with the functions $S_1$, $S_2$ defined in (\ref{S1_def}),
(\ref{S2_def}).

Now, one can expand the exponential function, retaining only the term linear in
$1/\beta J$:
\begin{eqnarray}\nonumber
\overline{G({\bf R})} = G_0({\bf R})\Bigg\{1 -2p - p \sum_{{\bf
r}\neq 0,{\bf R}} \Bigg(\frac{\pi}{8\beta J}F_1({\bf r},{\bf R})
\\\label{G2(R)_4}
+ \frac{\pi}{8\beta J(\pi-2)} F_2({\bf r},{\bf R}) \Bigg) + O((\beta J)^{-2})\Bigg\} .\quad
\end{eqnarray}
Then, using the asymptotic forms (\ref{Isc_asympt}) and (\ref{Iss_asympt}), and replacing the sum with an integral, one can show that, when $R = \sqrt{X^2 + Y^2} \to \infty$, the leading term comes from the integral which in polar coordinates reads as
\begin{eqnarray}\nonumber
&&\frac{1}{a^2} \int_{{\bf r}\neq 0, {\bf R}} d{\bf r} F_1({\bf r},{\bf R}) 
\\\nonumber
&&= \frac{R^2}{\pi^2} \int_{{\bf r}\neq 0,{\bf R}} \frac{rdrd\varphi}{r^2 (r^2 + R^2 - 2rR\cos\varphi)} + \ldots,
\end{eqnarray}
where the integral spans the entire system excluding areas
close to ${\bf r} = 0$ and ${\bf r} = \bf R$. This integration can be realized as follows:
\begin{eqnarray}\nonumber
\int_{{\bf r}\neq 0, {\bf R}} dr d\varphi &\to& \int_{a}^{R-a} dr \int_{0}^{2\pi} d\varphi + \int_{R+a}^{a\sqrt{N}} dr \int_{0}^{2\pi} d\varphi
\\\nonumber
&& + \int_{R-a}^{R+a} dr \int_{a/R}^{2\pi - a/R} d\varphi .
\end{eqnarray}

There is no difficulty in finding the integrals above, so, finally, one arrives at
\begin{eqnarray}\nonumber
\overline{G({\bf R})} = G_0({\bf R})\Big\{1 -2p - p
\frac{\pi}{2\pi\beta J}\ln(R/a) \Big\},
\end{eqnarray}
which can be written for small concentrations $p$ and low
temperatures $1/(\beta J)$ as
\begin{eqnarray}\label{G_s.d.(R)}
\overline{G({\bf R})} \simeq (1
-2p)\left(\frac{R}{a}\right)^{-\eta_{\mathrm{s.d.}}}
\end{eqnarray}
with 
\begin{eqnarray}\nonumber
\eta_{\mathrm{s.d.}}(p) &=& \eta(0) + p\frac{T}{2J} + O(p^2) + O((T/J)^2) 
\\\label{eta_s.d.}
&\simeq& \eta(0)(1+\pi p),
\end{eqnarray}
where $\eta(0)$ is the exponent of the pure system given by (\ref{eta_swa}). The factor $(1-2p)$ in (\ref{G_s.d.(R)}), which appeared naturally from the expansion, is the probability to have both sites that stand in the pair correlation function occupied with spins: $(1-p)^2 \underset{p\to 0}{\to} 1-2p$.

\section{\label{conclus}Conclusions}

\begin{figure}
\center{\includegraphics[width=0.35\textwidth,angle=-90]{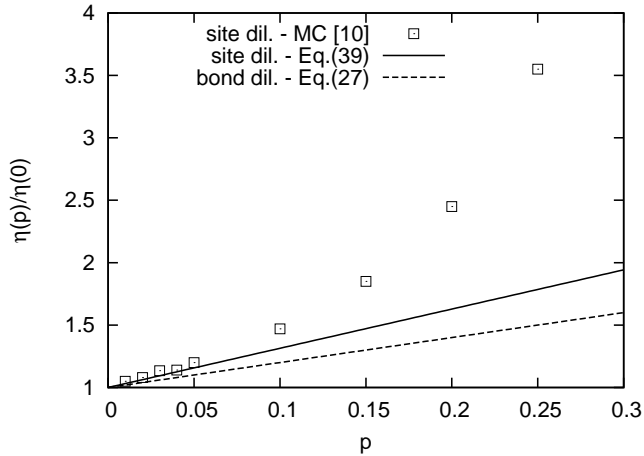}}
\caption{\label{fig2} Analytical (lines) and Monte Carlo (squares, site dilution only) results for the ratios $\eta_{\mathrm{s.d.}}(p)/\eta(0)$ and $\eta_{\mathrm{b.d.}}(p)/\eta(0)$ ($p$ is the concentration of missing spins and bonds, respectively). Concerning the analytical results one is referred to Eqs. (\ref{eta_s.d.}) and (\ref{eta_b.d.}). The Monte Carlo data are borrowed from \cite{KapikranianEtAl07} and come from simulations with Wolff cluster algorithm at $T/J=0.08$.}
\end{figure}

The spin-spin correlation function of the $2D$ $XY$ model decays as a power law at all temperatures below the Berezinskii-Kosterlitz-Thouless transition point with a temperature dependent exponent $\eta=\eta(T/J)$. In the $2D$ $XY$ model with site or bond dilution this exponent depends on concentration $p$ of removed sites/bonds as well. The knowledge of the slope $\partial\eta/\partial p$ at point $p=0$ allows to predict the value of the exponent for small dilution concentrations: $\eta(p)\simeq\eta(0)+p(\partial\eta/\partial p)|_{p=0}$. The analytical derivation, performed here in the low-temperature limit, led to $(\partial\eta/\partial p)|_{p=0}=\pi\eta(0)$ and $2\eta(0)$ for site and bond dilution, respectively, where $\eta(0)=T/2\pi J$ is the well known result for the model without dilution. These results are illustrated in Fig. \ref{fig2}.

The positive sign of $(\partial\eta/\partial p)|_{p=0}$ was well expected, since, as it was mentioned in Introduction, dilution can be interpreted as the increase of effective temperature. One might be tempted to equate the left sides of (\ref{eta_b.d.}) and (\ref{eta_s.d.}) to the universal value of $\eta(T_{\mathrm{BKT}})=1/4$ and identify the $T$ in the right side as the corresponding critical temperatures for site and bond dilution. Unfortunately, such an estimate of $T_{\mathrm{BKT}}(p)$ as a function of $p$ would not be quantitatively reasonable, since (\ref{eta_b.d.}), (\ref{eta_s.d.}) were obtained in the spin-wave approximation and do not hold for $T$ close to $T_{\mathrm{BKT}}(p)$.

It is worth noting that in order to compare the results for site and bond dilutions it may be more instructive to express the concentration of spinless sites, $p=$ (number of empty sites)/(number of all sites), through the actual concentration of missing bonds, $p'=$ ((four bonds)$\times$(number of empty sites))/(number of all bonds). (The latter relation holds, of course, only under the assumption of low dilution concentration, when the probability to have neighboring spinless sites is negligible.) Finally, noting that the total number of bonds in the system is two times the number of all sites, we have $p = p'/2$. Then one shall compare the exponent
\begin{equation}
\eta_{\mathrm{s.d.}}(p') = \eta(0) (1 + (\pi/2)p')
\end{equation}
and (\ref{eta_b.d.}) for $p'=p$, which means that we look at the systems with the same number of missing bonds (although in the case of site dilution all missing bonds are connected in unbreakable groups of four). One can notice that $\eta_{\mathrm{b.d.}} > \eta_{\mathrm{s.d.}}$ for the same concentration of missing bonds, which is well expected, since the disordering effect must be stronger for a completely random distribution of removed bonds in comparison to the site dilution case when removed bonds are connected in groups of four, and only these groups are distributed randomly then.

It also should be mentioned that, in principle, taking higher order terms in dilution concentration $p$ in (\ref{expansion_in_p}), one would expect to arrive at the end at the correlation function with exponent $\eta(p)$ represented by a series in powers of $p$ divergent at the percolation threshold value $p = p_{\mathrm{perc.}}$ for the square lattice [which is exactly $1/2$ for bond dilution and $\simeq 0.41$ for site dilution (see, for example, \cite{Newman_Ziff_2000})]. It is interesting in that it might give an exact value for the site percolation threshold which is not known yet. However, it might be as well not possible to carry out this calculation in an exact way, due to very high complexity.

\section{Acknowledgments}

I would like to thank Yurij Holovatch and Bertrand Berche, without whose guidance I would not start this problem in the first place, for useful discussions and corrections to the manuscript. I also acknowledge the support of the FP7 EU IRSES project N269139 'Dynamics and Cooperative Phenomena in Complex Physical and Biological Media' and the grant of the President of Ukraine for young scientists.

\appendix

\section{\label{AppendxA} Expression for $\left<\theta_{{\bf k}_1} \ldots \theta_{{\bf k}_{2n}} e^{i\sum_{\bf k} \eta_{\bf k} \theta_{\bf k}}\right>_0$}

Looking at (\ref{G(R)_four}), it is easy to see that
\begin{equation}
\left<\theta_{{\bf k}_1} \ldots \theta_{{\bf k}_{2n}}
e^{i\sum_{\bf k} \eta_{\bf k} \theta_{\bf k}}\right>_0 =
\frac{(-1)^n}{2^{2n}} \frac{\partial}{\partial\eta_{{\bf k}_1}}
\cdots \frac{\partial}{\partial\eta_{{\bf k}_{2n}}}\ G_0({\bf
R})\ ,
\end{equation}
where
\begin{equation}
\frac{\partial}{\partial\eta_{\bf k}} \equiv
\frac{\partial}{\partial\eta^c_{\bf k}} - i
\frac{\partial}{\partial\eta^s_{\bf k}}\ ,\quad
\frac{\partial}{\partial\eta_{\bf -k}} \equiv
\frac{\partial}{\partial\eta^c_{\bf k}} + i
\frac{\partial}{\partial\eta^s_{\bf k}}\ .
\end{equation}
Here and below, $\left<\dots\right>_0$ stands for the thermodynamic averaging with the Hamiltonian of the undiluted system, see Eq. (\ref{<...>_0}).

Noting that $\frac{\partial\eta_{\bf k}}{\partial\eta_{\bf
k'}} = 2 \delta_{\bf k,k'}$ ($\delta_{\bf k,k'}$ is Kronecker
delta) and establishing some simple recurrent relations when taking sequential derivatives from (\ref{G0(R)}), one relatively easy arrives at
\begin{eqnarray}\nonumber
\left<\theta_{{\bf k}_1} \ldots \theta_{{\bf k}_{2n}}e^{i\sum_{\bf
k} \eta_{\bf k} \theta_{\bf k}}\right>_0\ =\ G_0({\bf R})
\sum_{l=0}^{n} \frac{(-1)^{n-l}}{(2\beta J)^{2n-l}}
\\\label{theta...theta_e^eta_0} \times \sum_{\mathrm{comb.}(2n,l)}
\prod_{u=1}^{l} \frac{\delta_{{\bf k}_{i_u},-{\bf
k}_{j_u}}}{\gamma_{{\bf k}_{i_u}}} \prod_{w=1}^{2n-2l}
\frac{\eta_{-{\bf k}_{p_w}}}{\gamma_{{\bf k}_{p_w}}}\ ,\quad
\end{eqnarray}
where the sum $\displaystyle \sum_{\mathrm{comb.}(2n,l)}$
spans all distinguishable combinations of $l$ pairs $({\bf k}_{i_1},{\bf k}_{j_1})$, $({\bf k}_{i_2},{\bf k}_{j_2})$, ... $({\bf k}_{i_l},{\bf k}_{j_l})$ [combinations which can be obtained from each other by permutations of the pairs are not distinguished], which can be formed using ${\bf k}_1, {\bf k}_2,\ldots , {\bf k}_{2n}$. It is instructive to point out that
\begin{equation}\label{empty_sum_1}
\sum_{\mathrm{comb.}(2n,l)} 1 =
\frac{(2n)!}{(2!)^l(2n-2l)!l!} .
\end{equation}

Note, that when $\eta_{\bf k} = 0$, (\ref{theta...theta_e^eta_0}) gives
\begin{eqnarray}\label{theta...theta_0}
\left<\theta_{{\bf k}_1} \ldots \theta_{{\bf k}_{2n}}\right>_0
= \frac{1}{(2\beta J)^{n}} \sum_{\mathrm{comb.}(2n,n)}
\prod_{u=1}^{n} \frac{\delta_{{\bf k}_{i_u},-{\bf
k}_{j_u}}}{\gamma_{{\bf k}_{i_u}}} . \quad
\end{eqnarray}

\section{\label{AppendxB} Calculation of $\left< e^{i\sum_{\bf k} \eta_{\bf k} \theta_{\bf k}} H^n_1({\bf r}) \right>_0$}

To calculate the quantity 
\begin{eqnarray}\nonumber
\left< e^{i\sum_{\bf k} \eta_{\bf k} \theta_{\bf k}}
H^n_1({\bf r}) \right>_0 = (J/N)^n \sum_{{\bf k}_1,\ldots, {\bf k}_{2n}} e^{-i({\bf k}_1+ \ldots +{\bf k}_{2n}){\bf r}} 
\\\label{<exp(eta)(-beta H)>}
\times g_{{\bf k}_1,{\bf k}_2}\cdots g_{{\bf k}_{2n-1},{\bf k}_{2n}}\left<e^{i\sum_{\bf k} \eta_{\bf k} \theta_{\bf k}} \theta_{{\bf k}_1}\cdots \theta_{{\bf k}_{2n}}\right>_0 ,\qquad
\end{eqnarray}
[$g_{{\bf k},{\bf k}'}$ was defined after Eq. (\ref{H_1(r)_fourier})] one needs the result of the previous appendix for $\left<\theta_{{\bf k}_1} \ldots \theta_{{\bf k}_{2n}} e^{i\sum_{\bf k} \eta_{\bf k} \theta_{\bf k}}\right>_0$, Eq. (\ref{theta...theta_e^eta_0}). Each Kronecker delta from (\ref{theta...theta_e^eta_0}) deletes one summation index ${\bf k}'$ from the sum in (\ref{<exp(eta)(-beta H)>}) and ``connects" two ${\bf k}$'s belonging either to one $g$:
$$
N^{-1}\sum_{{\bf k},{\bf k}'} g_{{\bf k},{\bf k}'} \delta_{{\bf k},{\bf k}'}/\gamma_{\bf k} = N^{-1}\sum_{\bf k} g_{{\bf k},-{\bf k}}/\gamma_{\bf k},
$$ 
or to two different $g$'s:
$$
N^{-1}\sum_{{\bf k},{\bf k}'} g_{*,{\bf k}} g_{{\bf k}',*} \delta_{{\bf k},{\bf k}'}/\gamma_{\bf k} = N^{-1}\sum_{\bf k} g_{*,{\bf k}} g_{-{\bf k},*}/\gamma_{\bf k}.
$$
The former will be symbolically represented as $\overline{{}^\diagdown {g}^\diagup}$ and the latter as $g-g$. In the same vein, $g\times\eta$ will denote $N^{-1}\sum_{\bf k} g_{*,{\bf k}} \eta_{\bf k} e^{-i{\bf kr}}/\gamma_{\bf k}$. Note also, that $g_{{\bf k},{\bf k}'} = g_{{\bf k}',{\bf k}}$. 

Using (\ref{theta...theta_e^eta_0}) and the symbolic notations introduced above, one can write (\ref{<exp(eta)(-beta H)>}) as a sum of terms which are products of non-factorizable ``blocks" $\overline{{}^\diagdown {g}^\diagup}$, $\overline{{}^\diagdown {g-g}^\diagup}$, ..., $(\eta\times g\times\eta)$, $(\eta\times g-g\times\eta)$, ..., etc.:
\begin{eqnarray}\nonumber
\left< e^{i\sum_{\bf k} \eta_{\bf k} \theta_{\bf k}}
H^n_1({\bf r}) \right>_0 = G_0({\bf R}) \sum_{l=0}^{n} \frac{(-1)^{n-l}}{(2\beta J)^{2n-l}} \sum_{\mathrm{comb.}(2n,l)}  
\\\nonumber 
\times \left(\overline{{}^\diagdown {g}^\diagup}\right)^{\lambda_1} \left(\overline{{}^\diagdown {g-g}^\diagup}\right)^{\lambda_2}\cdots \Big(\overline{{}^\diagdown {\underbrace{g-\ldots-g}_{l}}^\diagup}\Big)^{\lambda_l}
\\\label{symbolic_sum}
\times (\eta\times g\times\eta)^{\lambda_1^*}\cdots (\eta\times\underbrace{g-\ldots-g}_{n}\times\eta)^{\lambda_n^*} . \qquad
\end{eqnarray}
To each term of the combinatorial sum $\sum_{\mathrm{comb.}(2n,l)}$, defined after (\ref{theta...theta_e^eta_0}), corresponds a certain set of integer numbers $\{\lambda_1, \ldots, \lambda_l, \lambda_1^*, \ldots \lambda_n^*\}$, $\lambda_i, \lambda_i^* = 0,1,2,\ldots$. However, there are many terms corresponding to the same set $\{\lambda_1, \ldots, \lambda_l, \lambda_1^*, \ldots \lambda_n^*\}$. Determining the number of terms (combinations of ``connections") in (\ref{symbolic_sum}) which correspond to any particular set of $\lambda$'s, one can use the $\lambda$'s as summation indexes. Using shorter notations
\begin{equation}\label{I_and_I*}
I_i = \overline{{}^\diagdown {\underbrace{g-g-\ldots-g}_{i}}^\diagup},\quad I_j^* = \eta\times\underbrace{g-g-\ldots-g}_{j}\times\eta
\end{equation}
[for explicite expressions for $I_i$, $I^*_i$ the reader is referred to (\ref{I_i_def})-(\ref{tilde_I_i})],
one arrives at
\begin{eqnarray}\nonumber
&&\left< e^{i\sum_{\bf k} \eta_{\bf k} \theta_{\bf k}}
H^n_1({\bf r}) \right>_0 = G_0({\bf R}) (2\beta)^{-n}
\sum_{l=0}^{n} \frac{(-1)^{n-l}}{(2\beta
J)^{n-l}} 
\\\nonumber
&&\times \left[\prod_{i=1}^{l}\sum_{\lambda_i = 0}^{\infty} \prod_{j=1}^{n}\sum_{\lambda^*_j = 0}^{\infty}\right] \delta\left(\sum_{i=1}^{l}i\lambda_i
 + \sum_{j=1}^{n}(j-1)\lambda^*_j -l\right)
\\\nonumber
&&\times \delta\left(\sum_{j=1}^{n}\lambda^*_j -(n-l)\right)\Lambda_{\lambda_1,\ldots,
\lambda_{l}}^{\lambda^*_1,\ldots,
\lambda^*_{n}} I_1^{\lambda_1}\cdots I_l^{\lambda_l}
 {I^*_1}^{\lambda^*_1}\cdots
{I^*_{n}}^{\lambda^*_{n}} ,
\\\label{I_avrg_*_1} 
\end{eqnarray}
where $\delta(x) = \Big\{ \begin{array}{ll} 1,\ x=0 \\ 0,\ x\neq 0
\end{array}$, and $\Lambda_{\lambda_1,\ldots, \lambda_{l}}^{\lambda^*_1,\ldots,
\lambda^*_{n}}$ is the combinatorial factor given by the number of combinations of connections in the sum in (\ref{symbolic_sum}) corresponding to the set $\{\lambda_1, \ldots, \lambda_l, \lambda_1^*, \ldots \lambda_n^*\}$. The upper possible values of $\lambda$'s are finite, of course, for finite $n$, but are not important (and so can be put equal to $\infty$ for simplicity), since the first Kronecker delta in (\ref{I_avrg_*_1}) assures that altogether one has $l$ connections between $g$'s and the second Kronecker delta assures that one has $(n-l)$ pairs of $\eta$'s; any realizations $\{\lambda_1, \ldots, \lambda_l, \lambda_1^*, \ldots \lambda_n^*\}$ that do not fulfill this conditions do not contribute to the sum.

Factor $\Lambda_{\lambda_1,\ldots, \lambda_{l}}^{\lambda^*_1,\ldots,
\lambda^*_{n}}$ can be found from a simple combinatorial analysis: it is given by the number of ways of dividing $n$ elements $g$ into $\lambda_1$ and $\lambda_1^*$ ``blocks" of {\it one} $g$, $\lambda_2$ and $\lambda_2^*$ ``blocks" of {\it two} $g$'s, and so on, which is given by (blocks with the same number of $g$'s are not distinguished)
\begin{eqnarray}\nonumber
&&n!/(\lambda_1! \lambda_2!\cdots\lambda_l!\ \lambda^*_1!
\lambda^*_2!\cdots\lambda^*_{n}!
\\\nonumber
&&\times (1!)^{\lambda_1} (2!)^{\lambda_2} \cdots (l!)^{\lambda_l} (1!)^{\lambda^*_1}(2!)^{\lambda^*_2} \cdots (n!)^{\lambda^*_{n}})\ ,
\end{eqnarray}
{\it times} the number of ways of connecting $g$'s inside every ``block". Consider a ``block" of $g_{{\bf k}_1,{\bf k}'_1}$, $g_{{\bf k}_2,{\bf k}'_2}$, ..., $g_{{\bf k}_i,{\bf k}'_i}$ and count in how many ways one can interconnect all $g$'s in it: $\overline{{}^\diagdown {\underbrace{g-g-\ldots-g}_{i}}^\diagup}$. The answer will be $2^{i-1} (i-1)!$, which is the number of permutations $i!$ divided by $2i$, since a) it is a cyclic structure (so only one $i$th part of all permutations give distinct combinations of interconnections, others are their repetitions) and b) the combination of connections is not changed by inversion of the $g$'s' order (hence only one half of the permutations must be counted), and multiplied by $2^i$, since every $g$ has {\it two} ${\bf k}$'s by which it can connect. The same reasoning leads to $2^{j-1}j!$ possible combinations of connections inside a ``block" $\eta\times\underbrace{g-g-\ldots-g}_{j}\times\eta$, since it is not cyclic. Eventually,
\begin{equation}\label{Lambda_1}
\Lambda_{\lambda_1,\ldots, \lambda_{l}}^{\lambda^*_1,\ldots,
\lambda^*_{n}} = n! \prod_{i=1}^{l} \frac{\left[2^{i-1}
(i-1)!\right]^{\lambda_i}}{\lambda_i! (i!)^{\lambda_i}}
\prod_{j=1}^{n} \frac{\left[2^{j-1}
j!\right]^{\lambda^*_j}}{\lambda^*_j! (j!)^{\lambda^*_j}}\ .
\end{equation}

Therefore, one has
\begin{eqnarray}\nonumber
&&\left< e^{i\sum_{\bf k} \eta_{\bf k} \theta_{\bf k}}
H^n_1({\bf r}) \right>_0 = G_0({\bf R}) \frac{n!}{\beta^n} 
\sum_{l=0}^{n} \frac{(-1)^{n-l}}{(2\beta J)^{n-l}}
\\\nonumber
&&\times \prod_{i=1}^{l}\sum_{\lambda_i = 0}^{\infty}
\frac{1}{\lambda_i!}\left( \frac{I_i}{2i} \right)^{\lambda_i}
\ \prod_{j=1}^{n}\sum_{\lambda^*_j = 0}^{\infty}
\frac{1}{\lambda^*_i!}\left( \frac{I^*_i}{2}
\right)^{\lambda^*_i}
\\\nonumber
&&\times \delta\left(\sum_{j=1}^{n} \lambda^*_j
-(n-l)\right) \delta\left(\sum_{i=1}^{l}i\lambda_i
 + \sum_{j=1}^{n}(j-1)\lambda^*_j -l\right).
\\\label{exp_1}
\end{eqnarray}
When $\eta_{\bf k} = 0$,
\begin{eqnarray}\nonumber
\left< H^n_1({\bf r}) \right>_0 &=&
(2\beta)^{-n} \left(\prod_{i=1}^{n}\sum_{\lambda_i = 0}^{\infty}\right) \delta\left(\sum_{i=1}^{n}i\lambda_i - n\right)
\\
&&\times \Lambda^{0,\ldots,0}_{\lambda_1,\ldots, \lambda_{n}} I_1^{\lambda_1}\cdots I_n^{\lambda_n}, \quad
\label{I_avrg_*_2}
\end{eqnarray}
and hence
\begin{equation}\label{exp_2}
\left< H^n_1({\bf r}) \right>_0 = \frac{n!}{\beta^n} \prod_{i=1}^{n}\sum_{\lambda_i = 0}^{\infty} \frac{1}{\lambda_i!}\left( \frac{I_i}{2i}
\right)^{\lambda_i} \delta\left(\sum_{i=1}^{n}i\lambda_i -n\right).
\end{equation}

\section{\label{AppendxC} Calculation of $I_i$ and $I^*_i$}

The sums $I_i$ and $I_i^*$, introduced in Appendix \ref{AppendxB}, (\ref{I_and_I*}), can be written as
\begin{equation}\label{I_i_def}
I_i = \frac{1}{N} \sum_{\bf k} \tilde{I}_{i-1} ({\bf k,-k})/\gamma_{\bf k}
\end{equation}
and
\begin{equation}\label{I*_i_def}
I^*_i = \frac{1}{N} \sum_{\bf k,k'} \tilde{I}_{i-1} ({\bf k,k}')
\frac{\eta_{-\bf k} \eta_{-\bf k'}}{\gamma_{\bf k}\gamma_{\bf k'}}\ e^{-i({\bf k
+ k'}){\bf r}^*}
\end{equation}
($i \ge 1$) with
\begin{equation}\label{tilde_I_i}
\tilde{I}_{i} ({\bf k,k}') = \frac{1}{N^{i}} \sum_{{\bf
k}_1,\ldots, {\bf k}_{i}} \frac{g_{{\bf k}, -{\bf k}_1} g_{{\bf k}_1,
-{\bf k}_2} \cdots g_{{\bf k}_{i-1}, -{\bf k}_{i}} g_{{\bf k}_{i},
{\bf k}'}}{\gamma_{{\bf k}_1} \cdots \gamma_{{\bf k}_i}}
\end{equation}
for $i \ge 1$ and $\tilde{I}_0({\bf k,k'}) = g_{\bf k,k'}$.
One can notice the obvious recurrent relation
\begin{equation}
\tilde{I}_{i+1} ({\bf k,k}') = \frac{1}{N} \sum_{{\bf k}^*}
\tilde{I}_{i} ({\bf k,-k}^*) g_{{\bf k}^*, {\bf k}'}/\gamma_{\bf k^*}\ .
\end{equation}

In the thermodynamic limit, one can replace the sum
$\frac{1}{N}\sum_{\bf k}$ over the 1st Brillouin zone with the
integral $\frac{a^2}{(2\pi)^2} \int_{-\pi/a}^{\pi/a} dk_x
\int_{-\pi/a}^{\pi/a} dk_y$, and then, noticing that
\begin{equation}\nonumber
\frac{a^2}{\pi^2}\int_{0}^{\pi/a} dk_x \int_{0}^{\pi/a} dk_y
\frac{\sin^4\frac{k_xa}{2}}{\sin^2\frac{k_xa}{2} +
\sin^2\frac{k_ya}{2}}\ =\ \frac{1}{\pi}
\end{equation}
and
\begin{eqnarray}\nonumber
\frac{a^2}{\pi^2}\int_{0}^{\pi/a} dk_x \int_{0}^{\pi/a} dk_y
\frac{\sin^2\frac{k_xa}{2}\cos^2\frac{k_xa}{2}}{\sin^2\frac{k_xa}{2}
+ \sin^2\frac{k_ya}{2}}\qquad\qquad\quad \\\nonumber =
\frac{a^2}{\pi^2}\int_{0}^{\pi/a} dk_x \int_{0}^{\pi/a} dk_y
\frac{\sin^2\frac{k_xa}{2}\sin^2\frac{k_ya}{2}}{\sin^2\frac{k_xa}{2}
+ \sin^2\frac{k_ya}{2}} =\ \frac{1}{2} - \frac{1}{\pi}\ ,
\end{eqnarray}
one can show that
\begin{eqnarray}\nonumber
\frac{1}{N}\sum_{\bf k'} g_{\bf k,-k'} g_{\bf k',k''}/ \gamma_{{\bf k}'} =
\left(1-\frac{2}{\pi}\right) g_{\bf k,-k''}
\\\nonumber
- \frac{1}{\pi} \left(g_{\bf k,-k''} + g_{\bf k,k''}\right) +
\left(\frac{1}{2}-\frac{1}{\pi}\right) \gamma_{\bf k}\gamma_{\bf k''}\ ,
\end{eqnarray}
\begin{eqnarray}\nonumber
\frac{1}{N}\sum_{\bf k'} g_{\bf k,k'} g_{\bf k',k''}/ \gamma_{{\bf k}'} =
\left(1-\frac{2}{\pi}\right) g_{\bf k,k''}
\\\nonumber
- \frac{1}{\pi} \left(g_{\bf k,-k''} + g_{\bf k,k''}\right) +
\left(\frac{1}{2}-\frac{1}{\pi}\right) \gamma_{\bf k}\gamma_{\bf k''}\ ,
\end{eqnarray}
and
\begin{equation}\nonumber
\frac{1}{N}\sum_{\bf k} g_{\bf k,k'} = -
\gamma_{\bf k'} .
\end{equation}
Then, it is easy to see that
\begin{eqnarray}\nonumber
\tilde{I}_{i} ({\bf k,k}') = A_i g_{{\bf k},(-1)^i{\bf k}'} + B_i
\left(g_{\bf k,-k'} + g_{\bf k,k'}\right) + C_i \gamma_{\bf k}\gamma_{\bf k'}
\end{eqnarray}
with coefficients $A_i$, $B_i$ and $C_i$ obeying the recursive
relations
\begin{equation}\nonumber
A_{i+1} = \left(1-\frac{2}{\pi}\right) A_i\ ,
\end{equation}
\begin{equation}\nonumber
B_{i+1} = -\frac{1}{\pi} A_i + \left(1-\frac{4}{\pi}\right) B_i\ ,
\end{equation}
\begin{equation}\nonumber
C_{i+1} = \left(\frac{1}{2}-\frac{1}{\pi}\right) \left(A_i + 2
B_i\right) - C_i\ ,
\end{equation}
and $A_0 = 1$, $B_0 = 0$, $C_0 = 0$. Thus,
\begin{equation}\nonumber
A_{i} = \left(1-\frac{2}{\pi}\right)^i\ ,
\end{equation}
\begin{eqnarray}\nonumber
B_{i} &=& -\frac{1}{\pi} \sum_{j=0}^{i-1}
\left(1-\frac{4}{\pi}\right)^{j}
\left(1-\frac{2}{\pi}\right)^{i-1-j} \\\nonumber &=&
-\frac{1}{2}\left[\left(1-\frac{2}{\pi}\right)^{i} -
\left(1-\frac{4}{\pi}\right)^{i}\right]\ ,
\end{eqnarray}
\begin{eqnarray}\nonumber
C_{i} &=& (-1)^{i-1}\left(\frac{1}{2}-\frac{1}{\pi}\right)
\sum_{j=0}^{i-1} (-1)^{j} \left(1-\frac{4}{\pi}\right)^{j}
\\\nonumber
&=& \frac{1}{4} \left[ (-1)^{i-1} +
\left(1-\frac{4}{\pi}\right)^{i} \right]\ .
\end{eqnarray}

Finally, one can obtain expressions for $I_i$ and $I^*_i$ and find
that
\begin{eqnarray}\nonumber
&& \sum_{i=1}^{\infty} (-1)^i I^*_i 
\\\nonumber
&&= \left\{\begin{array}{lll} \frac{1}{N} {\displaystyle \sum_{\bf k,k'}} \Big(\left[\frac{\pi}{4} - \frac{\pi}{4(\pi-2)}\right] g_{\bf k,-k'} - \left[\frac{\pi}{4} + \frac{\pi}{4(\pi-2)}\right] g_{\bf k,k'}\Big) 
\\ 
\times \frac{\eta_{-\bf k}\eta_{-\bf k'}}{\gamma_{\bf k}\gamma_{\bf k'}} e^{-i({\bf k+k'}){\bf r}},\ \textrm{if}\ {\displaystyle\sum_{\bf k,k'}} \eta_{\bf -k} \eta_{\bf -k'} e^{-i({\bf k+k'}){\bf r}} = 0;
\\ 
\infty,\quad \textrm{otherwise}. \end{array} \right.
\\\label{sum_I*_i}
\end{eqnarray}

\section{\label{AppendxD} Functions $S_{1}(A,B)$ and $S_{2}(A,B)$}

In this appendix one finds the asymptotic form for the functions
\begin{equation}\label{S1_def}
S_{1}(A,B) = \frac{1}{N} \sum_{\bf k}
\frac{\sin\frac{k_xa}{2}\cos\frac{k_xa}{2}}{\sum_{\alpha=x,y}\sin^2\frac{k_\alpha
a}{2}} \sin Ak_x \cos Bk_y\ ,
\end{equation}
\begin{equation}\label{S2_def}
S_{2}(A,B) = \frac{1}{N} \sum_{\bf k}
\frac{\sin^2\frac{k_xa}{2}}{\sum_{\alpha=x,y}\sin^2\frac{k_\alpha
a}{2}} \cos Ak_x \cos Bk_y\ ,
\end{equation}
where the sums span the 1st Brillouin zone. It turns out that
simple analytic expressions can be obtained, assuming that at
least one of the arguments $A,B$ is large. Using the integral
\cite{Prudnikov}
\begin{equation}
\int_{0}^{\infty} \frac{\cos x}{x^2 + a^2} dx = \frac{\pi}{2|a|}
e^{-|a|} ,
\end{equation}
one can show that
\begin{eqnarray}\nonumber
S_{1}(A\to\infty,B) = \frac{a}{\pi} \int_{0}^{\pi/a} dk_y
e^{-A\frac{2}{a}\sin\frac{k_ya}{2}} \cos Bk_y
\\\nonumber
\times
\frac{\sinh\left(2\sin\frac{k_ya}{2}\right)}{2\sin\frac{k_ya}{2}}
\simeq \frac{a}{\pi} \int_{0}^{\pi/a} dk_y e^{-A k_y}\cos B k_y
\end{eqnarray}
and
\begin{eqnarray}\nonumber
S_{1}(A,B\to\infty) = \frac{a}{\pi} \int_{0}^{\pi/a} dk_x
e^{-B\frac{2}{a}\sin\frac{k_xa}{2}} \sin Ak_x
\\\nonumber
\times \cos\frac{k_xa}{2} \simeq \frac{a}{\pi} \int_{0}^{\pi/a} dk_y
e^{-B k_x}\sin Ak_x
\end{eqnarray}
So,
\begin{equation}\label{Isc_asympt}
S_{1}(A,B) = \frac{a}{\pi} \frac{A}{A^2 + B^2},
\end{equation}
when at least one of its arguments $A,B$ is sufficiently large.

In a similar way one can show that
\begin{equation}\label{Iss_asympt}
S_{2}(A,B) = \frac{a^2}{2\pi} \frac{B^2 - A^2}{\left(A^2 +
B^2\right)^2},
\end{equation}
if at least one of its arguments $A,B$ is sufficiently large.



\begin{thebibliography}{99}
\bibitem{footnote} Comparing to \cite{KapikranianEtAl07}, where only the first three orders of the expansion in powers of $H(\{p_{\bf r}\})$ [see (\ref{H0+H1})] where taken into account, here, the entire infinite series is considered. It has to be admitted also that the third order contribution was estimated incorrectly in \cite{KapikranianEtAl07}!
\end{thebibliography}
\end{document}